\documentclass[pdflatex,sn-mathphys-ay]{sn-jnl}

\usepackage{graphicx}%
\usepackage{multirow}%
\usepackage{amsmath,amssymb,amsfonts}%
\usepackage{amsthm}%
\usepackage{mathrsfs}%
\usepackage[title]{appendix}%
\usepackage{xcolor}%
\usepackage{textcomp}%
\usepackage{manyfoot}%
\usepackage{booktabs}%
\usepackage{algorithm}%
\usepackage{algorithmicx}%
\usepackage{algpseudocode}%
\usepackage{listings}%

\theoremstyle{thmstyleone}%
\theoremstyle{thmstyletwo}%

\theoremstyle{thmstylethree}%

\begin{document}

\title[Methodological refinement of the submillimeter galaxy cross-correlation]{Methodological refinement of the submillimeter galaxy cross-correlation function measurements and their uncertainty estimation}


\author*[1,2]{\fnm{Joaqu{\'i}n} \sur{Gonz{\'a}lez-Nuevo}}\email{gnuevo@uniovi.es}

\author[1,2]{\fnm{Laura} \sur{Bonavera}}\email{bonaveralaura@uniovi.es}
\equalcont{These authors contributed equally to this work.}

\author[3,4,1,2]{\fnm{Marcos} \spfx{M.} \sur{Cueli}}\email{mcueli@uniovi.es}
\equalcont{These authors contributed equally to this work.}

\author[1,2]{\fnm{David} \sur{Crespo}}\email{crespodavid@uniovi.es}
\equalcont{These authors contributed equally to this work.}

\author[1,2,5,6]{\fnm{Jose Manuel} \sur{Casas}}\email{casasjm@uniovi.es}
\equalcont{These authors contributed equally to this work.}

\author[1,2]{\fnm{Rebeca} \sur{Fern{\'a}ndez-Fern{\'a}ndez}}\email{fernandezferrebeca@uniovi.es}
\equalcont{These authors contributed equally to this work.}

\affil*[1]{\orgdiv{Departamento de F{\'i}sica}, \orgname{Universidad de Oviedo}, \orgaddress{\street{C. Federico Garcia Lorca 18}, \postcode{33007}, \state{Oviedo}, \country{Spain}}}

\affil[2]{\orgname{Instituto Universitario de Ciencias y Tecnolog{\'i}as Espaciales de Asturias (ICTEA)}, \orgaddress{\street{C. Independencia 13}, \postcode{33004}, \city{Oviedo}, \country{Spain}}}

\affil[3]{\orgname{SISSA}, \orgaddress{\street{Via Bonomea 265},  \postcode{34136}, \city{Trieste}, \country{Italy}}}

\affil[4]{\orgname{IFPU - Institute for fundamental physics of the Universe}, \orgaddress{\street{Via Beirut 2},  \postcode{34014}, \city{Trieste}, \country{Italy}}}

\affil[5]{\orgname{Instituto de Astrof{\'i}sica de Canarias}, \orgaddress{\postcode{38205}, \city{La Laguna}, \country{Spain}}}

\affil[6]{\orgdiv{Departamento de astrof{\'i}sica}, \orgname{Universidad de la Laguna}, \orgaddress{\postcode{38206}, \city{La Laguna}, \country{Spain}}}




\abstract{\textbf{Purpose:} In this study, we aim to develop a new methodology to estimate the cross-correlation function and uncertainties and apply it to the analysis of magnification bias in galaxy surveys.
 
\textbf{Methods:} We adopt a new methodological framework that uses a statistically rigorous approach to obtain more robust measurements for constraining cosmological parameters. This strategy involves using the full field area to count the number of different pairs for each field and combine them into a single estimation, reducing statistical uncertainty and accounting for the full information available in the data. The covariance matrix was estimated internally using an oversampled bootstrap method. We divided each field into at least five patches, that were defined automatically using a k-mean clustering algorithm. 

\textbf{Results:} We investigate the robustness of the new methodology by comparing the results from a spectroscopic lens sample with those from a photometric lens sample, finding them to be compatible. We also analyse the cross-correlation function and auto-correlation function for individual fields in the three GAMA fields, comparing both samples. The G15 field was found to have a stronger signal compared to the other fields, suggesting that the stronger cross-correlation is produced by the rare combination of two excesses of large-scale structure in both the foreground and background samples.

\textbf{Conclusion:} Our results demonstrate the robustness of the new methodology and suggest that the differences with respect to the mini-tile previously used approached may be due to physical properties of the samples themselves. The identified G15 anomalous signal warrants further investigation into its impact on cosmological parameters.}


\maketitle

\section{Introduction}
 Magnification bias, the excess number of high-redshift sources near low-redshift mass structures, is a promising cosmological probe that can shed light on the large-scale structure of the universe and the distribution of dark matter \citep[see][and references therein]{CUE25}. Magnification bias measurements complement other cosmological probes and provide independent constraints on cosmological parameters. This phenomenon is caused by gravitational lensing, where the distribution of matter in the universe, such as galaxies or clusters of galaxies, bends light as it travels through space, magnifying or distorting the apparent brightness and size of distant objects \citep[see e.g.,][]{SCH92}.

 This phenomenon can be clearly seen when two source samples that do not share the same redshift range feature a significant angular cross-correlation function. The only possible explanation for this is that the foreground sources are magnifying the flux of background sources through gravitational lensing. Several studies have detected this effect in different scenarios: for example, between galaxies and quasars \citep{SCR05, MEN10}, between Herschel sources and Lyman-break galaxies \citep{HIL13}, or between the cosmic microwave background (CMB) and other sources \citep{Bia15,Bia16}.
 
In this paper, we focus on sub-millimeter galaxies (SMGs) as background sources due to their optimal properties for lensing studies, such as their steep luminosity function, high redshifts, and faint emission in the optical band \citep[see, for example,][among the most important ones]{Bla96,Neg07,Neg10,Neg17,GON12,GON17,Bus12,Bus13,Fu12,War13,CAL14,NAY16,Bak18,Bak20}. Several publications have demonstrated the magnification bias effect on SMGs, and it has been measured with high significance \citep{GON14,GON17,GON21,BON19,BON20,CUE21,FER22,Cre22, Cue23}. Moreover, the ability to split the foreground sample into different redshift bins allows for a more detailed tomographic analysis \citep{BON21, BON23,CUE22}.

It is worth noting that all these previous results are based on the common area between the Herschel Astrophysical Terahertz Large Area Survey \citep[H-ATLAS;][]{EAL10} and the Galaxy and Mass Assembly \citep[GAMA;][]{DRI11,BAL10,BAL14,LIS15} surveys. This common area is divided into four independent fields: GAMA09 (G09; 48.4 square degrees), GAMA12 (G12; 48.6 square degrees), GAMA15 (G15; 49.4 square degrees), and south Galactic pole (SGP; 60 square degrees). Although these fields are technically large enough to derive cosmological constraints, they are still small compared to other wide-area galactic surveys such as the Sloan Digital Sky Survey \citep[SDSS; ][]{BLA17,AHU19}, the Dark Energy Survey \citep[DES;][]{DES05}, or the future \textit{Euclid} \citep{EuclidI_22}. Therefore, there is a possibility that these previous results were affected by sample size limitations or sample variance statistical issues. In this field, the former occurs when the number of galaxies in the sample is not large enough to draw significant conclusions for the whole population. Sample variance, on the other hand, quantifies deviations among the values of an estimator across different sky regions.

To minimise these effects, the most effective strategy is to increase the overall area. This is certainly possible, but there are caveats that make the transition not straightforward. For example, H-ATLAS has an additional field, the north Galactic pole or NGP, with 199.6 square degrees, but it was not covered by the GAMA survey. Several other independent surveys, such as the \textit{Herschel} Multi-tiered Extragalactic Survey \citep[HerMES;][]{Oli12}, or the Great Observatories Origins Deep Survey-\textit{Herschel}, \citep[GOODS-H;][]{Elb11}, have observed wide fields with Herschel, but each of them has slightly different characteristics that need to be homogenised before being combined into a single analysis. Additionally, there is probably no single foreground survey that covers all of them. In any case, we are planning to move in that direction in the near future. However, before addressing such challenges, we aim to review and expand on the methodology used in our previous studies with well-known and established data before increasing the available area by analysing all the available Herschel wide area surveys. 

This work concentrates on the measurement itself of the angular cross-correlation function induced by magnification bias and its uncertainties. This aspect is of fundamental importance to accurately assess the statistical significance of the magnification bias signal in the observational data. The paper provides a detailed description of a new methodology used to measure the cross-correlation function, along with an in-depth analysis of the sources of uncertainty. It also proposes strategies to mitigate these uncertainties and improve the accuracy of the measurements. This new methodology was already successfully applied in \citet{Cue23}, which assesses cosmological parameter constraints with a single redshift bin, and \citet{BON23}, that builds upon this framework by extending the analysis to a tomographic setup.

The paper is structured as follows: Sect. \ref{sec:data} describes the data used in the analysis. Sect. \ref{sec:method} details the measurement methodology employed to quantify the magnification bias effect. The estimation of uncertainties is discussed in Sect. \ref{sec:COV}. Sect. \ref{sec:robust} investigates the robustness of the measurements and tests their potential sample variance. Finally, Sect. \ref{sec:concl} summarises the main conclusions.

\section{Data}
\label{sec:data}
\subsection{Background sample}
H-ATLAS is the largest extragalactic survey conducted by the Herschel space observatory \citep{PIL10}, covering an extensive area of approximately 610 square degrees. The Photodetector Array Camera and Spectrometer \citep[PACS;][]{POG10} and the Spectral and Photometric Imaging Receiver \citep[SPIRE;][]{GRI10} instruments were used in the survey, which spanned the wavelength range between $100\,\mu m$ and $500\,\mu m$. Detailed information on the H-ATLAS map-making, source extraction, and catalogue generation can be found in \citet{IBA10}, \citet{PAS11}, \citet{RIG11}, \citet{VAL16}, \citet{BOU16}, and \citet{MAD20}.

The background sample for this study was derived from H-ATLAS sources detected in the three GAMA survey equatorial fields (total common area of $\sim147$ deg$^2$, G09, G12 and G15), the North Galactic Pole (NGP) field, and the part of the South Galactic Pole (SGP) field that overlaps with the spectroscopic foreground sample. We applied a photometric redshift selection of $1.2<z<4.0$ to avoid any overlap in the redshift distributions of lenses and background sources, leaving us with $\sim 66,000$ sources ($\sim24$\% of the initial sample and $\langle z_{\mathrm{ph}}\rangle=2.20$). The redshift estimation is described in detail in \citet{GON17} and \citet{BON19}. This background sample is identical to the one used in previous works \citep{GON14,GON17,GON21,BON19,BON20,BON21,CUE21,CUE22,FER22,Cre22}.

\begin{figure}[ht]
\includegraphics[width=\columnwidth]{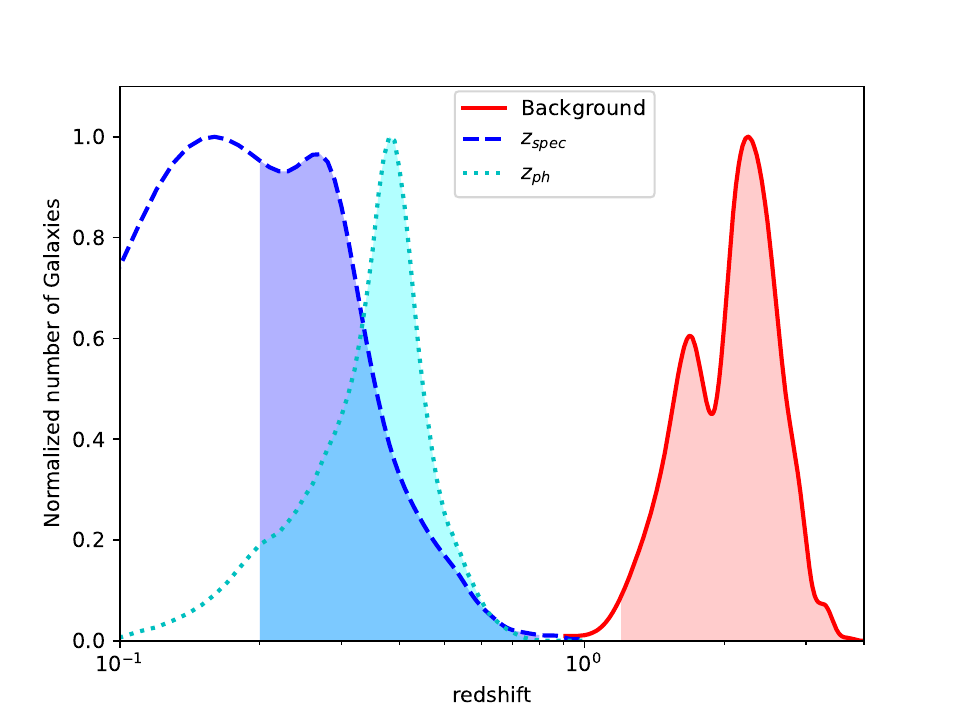}
 \caption{Normalised redshift distributions of the three catalogues used in this work: the background sample, that is, H-ATLAS high-z SMGs (solid red line); the GAMA spectroscopic foreground sample (dashed dark blue line); and the SDSS photometric foreground sample (dotted light blue line).
 }
 \label{fig:zhist}
\end{figure}

The redshift distribution of the background sources used in our analysis (represented by the red line in Fig. \ref{fig:zhist}) is estimated as $p(z|W)$ for galaxies selected by our window function, which is a top hat function for the redshift range of $1.2 < z < 4.0$. This distribution takes into account the effect of random errors in photometric redshifts, following the methodology described in \citet{GON17}. In that work, it is demonstrated that the potential contamination of lower-redshift sources ($z < 0.8$) with photometric redshifts above 1.2 is statistically negligible, even when accounting for the uncertainties in photometric redshift estimation \citep[see][and references therein]{GON17}.

\subsection{Foreground samples}
This study employs two distinct foreground samples, each selected independently. The first sample is the default one employed in previous works \citep{GON21,BON20,BON21,CUE21,CUE22} and is referred to as the "z$_{spec}$ sample". This sample comprises approximately 150,000 galaxies in the range of $0.2 < z_{spec} < 0.8$ (with a median value of $z_{spec}$ = 0.28; see dark blue shadowed region in Fig. \ref{fig:zhist}), obtained from the GAMA II spectroscopic survey \citep{DRI11,BAL10,BAL14,LIS15}. The H-ATLAS and GAMA II surveys were coordinated to maximise their overlap, covering the three equatorial regions and the SGP one. The total common area used in this study was approximately 207 square degrees, considering only the regions where both surveys overlap.

The second foreground sample was obtained from the 16th data release of the Sloan Digital Sky Survey \citep[SDSS;][]{BLA17,AHU19}, consisting of galaxies with photometric redshift within the range of $0.2 < z_{ph} < 0.8$ and photometric redshift error $z_{err}/(1+z) < 1$ (photoErrorClass = 1). This sample, called the "z$_{ph}$ sample," covers a total area of approximately 317 square degrees, encompassing the H-ATLAS equatorial regions and the NGP, and includes roughly 962,000 galaxies in the common area, with a median value of $z_{ph}$ = 0.38 (see light blue shadowed region in Fig. \ref{fig:zhist}). The second sample was introduced to investigate the impact of increasing the density of potential lenses on the measurements and their robustness with respect to the use of different foreground samples.

\section{Methodological aspects}
\label{sec:method}

\subsection{Tiling strategy}
\label{sec:tiling}
\begin{figure*}[ht]
\includegraphics[width=\textwidth]{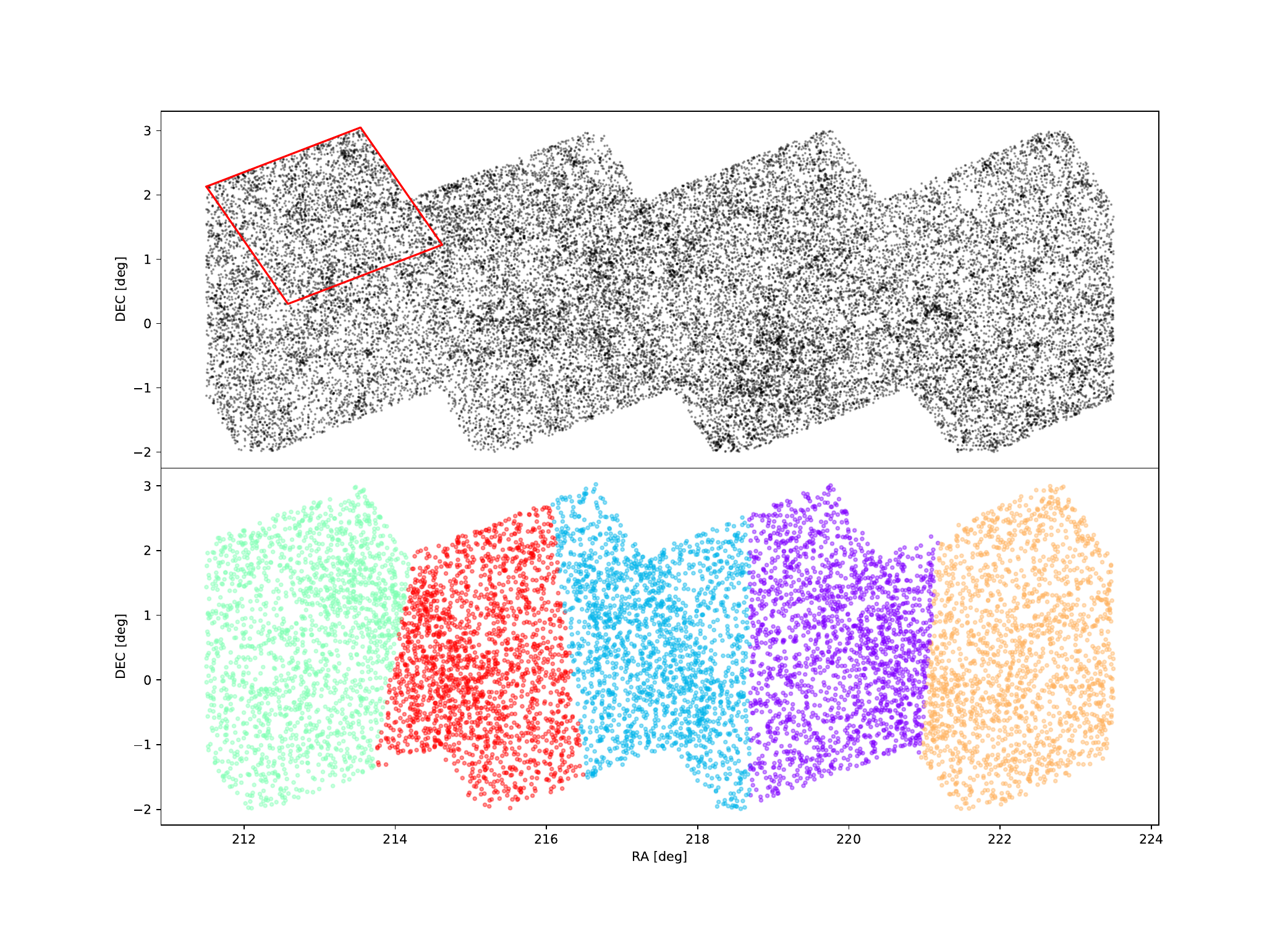}
 \caption{The spatial distribution of foreground (top panel) and background (bottom panel) sample galaxies for the G15 field. For the background sample, different colours indicate patch definitions, with five of them shown (approximately 9.6 square degrees each). The red square in the top panel indicates the typical shape and size of a ``mini-tile'' (see text for more details).
 }
 \label{fig:patches5}
\end{figure*}

\begin{figure*}[ht]
\includegraphics[width=\textwidth]{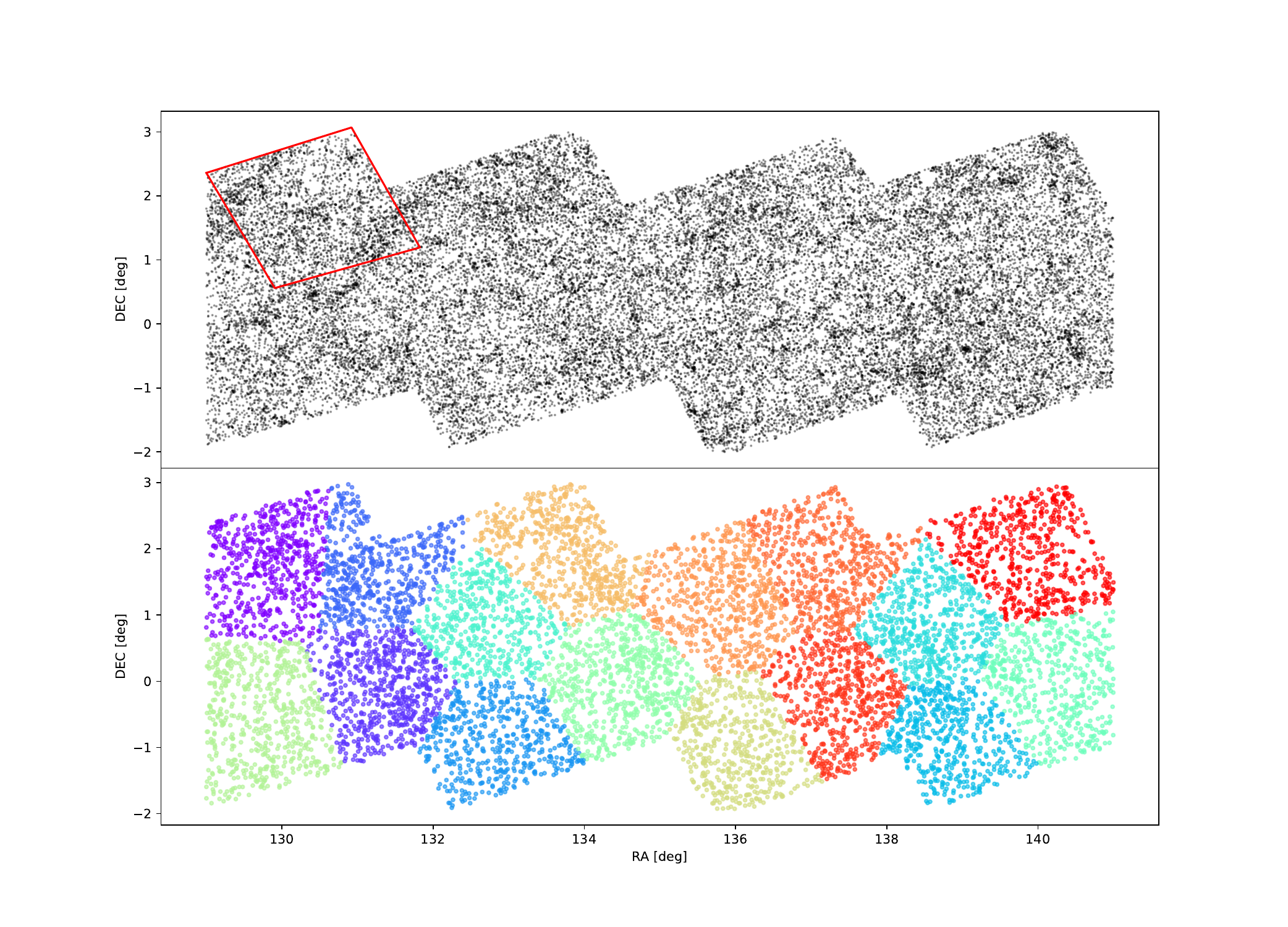}
 \caption{The spatial distribution of foreground (top panel) and background (bottom panel) sample galaxies for the G09 field. For the background sample, different colours indicate patch definitions, with 16 of them shown (approximately 3 square degrees each). The red square in the top panel indicates the typical shape and size of a ``mini-tile'' (see text for more details).
 }
 \label{fig:patches16}
\end{figure*}

The H-ATLAS survey used a scanning strategy that resulted in overlapping rhomboidal shapes, or tiles, in most fields, each covering an area of approximately 16 square degrees. However, as outlined in the detailed analysis by \citet{GON21}, dividing the full area in smaller ``mini-tiles'' produces more robust cross-correlation measurements with fewer required corrections. Mini-tiles are obtained by dividing a tile into four equal parts, each covering approximately 4 square degrees, which are then adapted to the field limits. As the size of the mini-tiles is similar to the angular distances we would like to study, the measurements need to be corrected for the so-called Integral Constraint (IC), which we discuss in detail in Sect. \ref{sec:IC}. Examples of mini-tile shapes and sizes are shown by the red squares in Figs. \ref{fig:patches5} and \ref{fig:patches16}. The use of mini-tiles was also the default tiling scheme in previous related works, such as \citet{BON20,BON21} and \citet{CUE21,CUE22}.

In this work, we revisit the conclusions of \citet{GON21} regarding measurement strategies for the H-ATLAS fields. According to their study, the mean value of the measurements taken using the full area of each field have smaller uncertainties than averaging the signal over smaller patches like mini-tiles and do not require an IC correction. However, this approach was disregarded due to an additional large angular scale bias that is difficult to correct. Because of its advantages, in this study we explore the possibility of using the combined available area for a single cross-correlation measurement and investigate the nature and impact of the large angular scale bias. A statistical characterization of the three estimation approaches considered in this work and a justification for this choice are provided in Appendix \ref{secA1}.

Even if the mini-tile scheme is not used for the measurement itself, a subdivision of all the available area into minimal subregions is still necessary to assign meaningful uncertainties to the measurement and, more generally, to estimate the covariance matrix internally, as discussed in Sect. \ref{sec:COV}. However, using mini-tiles poses a practical challenge as the vertices of each of them must be manually defined. While this task was accomplished for the four H-ATLAS fields with 63 mini-tiles, it becomes increasingly time-consuming when expanding the study to other wide field surveys observed by Herschel, which could have hundreds of mini-tiles. To address this issue, we have adopted a more flexible subregion definition procedure that maintains similar properties to the previous definition but is more scalable for future studies.

To define the subregions, we drew inspiration from the methodology used by TreeCorr, a popular software package for measuring galaxy clustering \citep{TreeCorr}. TreeCorr uses a k-means clustering algorithm to partition the data into subregions known as ``patches'', which are similar in size and shape to our mini-tiles. Specifically, we adopted the k-means algorithm provided by the SciPy library. The algorithm aims to minimise the sum of squared distances between data points and their assigned centroid. To determine the number of clusters (i.e., the patches), we imposed a minimum area for each cluster. We added an additional step by repeating the procedure 10 times with different random initial centroids, and selecting the run that yielded the most consistent number of data points across different clusters.

Figures \ref{fig:patches5} and \ref{fig:patches16} show the distribution of foreground (top panels) and background (bottom panels) sample galaxies for the G15 and G09 fields, respectively. Different colours are used to indicate examples of patch definitions, with five clusters (approximately 9.6 square degrees each) shown for the G15 and 16 clusters (approximately 3 square degrees each) shown for the G09. The particular number of clusters is related to the estimation of the covariance matrix and will be discussed in Sect. \ref{sec:COV}. Due to the specific limits of the field, clusters with smaller areas tend to adopt more regular shapes.

\subsection{Auto/Cross-correlation estimators}
The angular auto-correlation function $w_{\text{auto}}(\theta)$ is estimated with the Landy-Szalay estimator \citep[][]{LAN93}, 
\begin{equation}
    \label{eq:LS_est}
    \hat{w}_{\text{auto}}(\theta) = \dfrac{\text{DD}(\theta) - 2 \text{DR}(\theta) + \text{RR}(\theta)}{RR(\theta)},
\end{equation}
where $\text{DD}$, $\text{DR}$ and $\text{RR}$ are the normalised data-data, data-random and random-random pair counts for a given angular separation $\theta$.

On the other hand, as described in detail in \citet{GON17} and \citet{BON20}, the cross-correlation function measurement is performed via a modified version of the \cite{LAN93} estimator \citep{HER01},
\begin{equation}
\hat{w}_{\text{cross}}(\theta)=\frac{\rm{D}_f\rm{D}_b(\theta)-\rm{D}_f\rm{R}_b(\theta)-\rm{D}_b\rm{R}_f(\theta)+\rm{R}_f\rm{R}_b(\theta)}{\rm{R}_f\rm{R}_b(\theta)},
\end{equation}
where $\rm{D}_f\rm{D}_b$, $\rm{D}_f\rm{R}_b$, $\rm{D}_b\rm{R}_f$ and $\rm{R}_f\rm{R}_b$ are the normalised foreground-background, foreground-random, background-random and random-random pair counts for a given angular separation $\theta$.

The instrumental noise caused by the scanning strategy leads to a surface density variation in the background galaxy sample \citep[see Fig. \ref{fig:patches5} and more clearly in Fig. 4, top panel, of][]{GON21}. The overlap between the tiles reduces such noise and allows fainter sources to be detected. To correct for this bias, we generated random catalogues using the same procedure as in \citet{AMV19}: a flux is randomly chosen using the cumulative probability distribution of fluxes of the real sources. Then, a random position on the image is generated, and the local noise is estimated as the quadratic sum of the instrumental noise in that pixel and the confusion noise. The source is kept if its flux, perturbed by a Gaussian deviation equal to the total local noise estimate, was greater than $4\sigma$; otherwise, the process is repeated starting from choosing a flux randomly. 

After accounting for the instrumental noise variation, we observed a slight decrease in the cross-correlation function at the largest angular scales. This kind of correction is not needed for the foreground galaxy samples.

In the case of the mini-tiles approach, as in previous works, we compute the angular (auto-)cross-correlation function for each mini-tile. To ensure stability, we average over ten different random catalogue realisations. By averaging over all minitiles the value of the cross-correlation functions is then estimated for a given angular separation bin. The uncertainties are computed as the standard error of the mean, $\sigma_{\mu}=\sigma/\sqrt{n}$, where $\sigma$ is the standard deviation of the population and $n$ is the number of independent areas. Each selected region is assumed to be statistically independent due to the minimal overlap. 

In \citet{GON21}, the authors also compute the mean value of the four cross-correlation functions estimated in each entire field area and the standard error of the mean as its uncertainty. In this work, we adopted a more statistically rigorous approach by counting the number of different foreground-background pairs for each field and combining them into a single cross-correlation estimation. This method accounts for the full information available in the data and reduces the statistical uncertainty inherent to measuring four separate (auto-)cross-correlation functions. As demonstrated in Appendix \ref{secA1}, this approach yields systematically smaller uncertainties and lower bias compared to both the mini-tile and zone-averaging alternatives. However, when more independent fields become available in the future, the two approaches are expected to converge to similar results due to the increased statistical power and reduced sample variance. The uncertainties associated to this approach will be discussed in detail in Sect. \ref{sec:COV}.

\subsection{Discussion on the IC}
\label{sec:IC}

When averaging over multiple regions, the observed correlation function can be weakened at scales close to the size of the region due to large-scale fluctuations. This leads to a bias in the estimated (auto-) cross-correlation function \citep{Infante94, RE99}, , where the ``true'' measurement, $\hat{w}_{\text{ideal}}(\theta)$ is related to the observed measurement $\hat{w}(\theta)$ by a constant that is precisely the IC, i.e.
\begin{equation}
\label{eq:ic}
    \hat{w}_{\text{ideal}}(\theta) = \hat{w}(\theta) + \text{IC}.
\end{equation}

There are theoretical approaches to estimate the IC for a particular scanning strategy \citep[see e.g.,][]{Ade05}, but in practice, it is commonly estimated numerically using RR counts. Specifically, the IC can be estimated for the cross-correlation using the formula:

\begin{equation}
\text{IC}=\frac{\sum_i{\rm{R}_{\text{f}}\rm{R}_{\text{b}}(\theta_i)w_{\text{ideal}}(\theta_i)}}{\sum_i{\rm{R}_{\text{f}}\rm{R}_{\text{b}}(\theta_i)}},
\end{equation}
where $w_{\text{ideal}}(\theta)$ is an assumed model for the cross-correlation function. An equivalent expression is used for the auto-correlation.

In \citet{GON21}, a power-law model was assumed $w_{\text{ideal}}(\theta)= A \theta^{-\gamma}$, where $A$ and $\gamma$ were the best-fit parameters obtained by fitting only the observed cross-correlation function below 20 arcmin. For the mini-tiles, the derived IC value was $9\times10^{-4}$.

\begin{figure}[ht]
\includegraphics[width=\columnwidth]{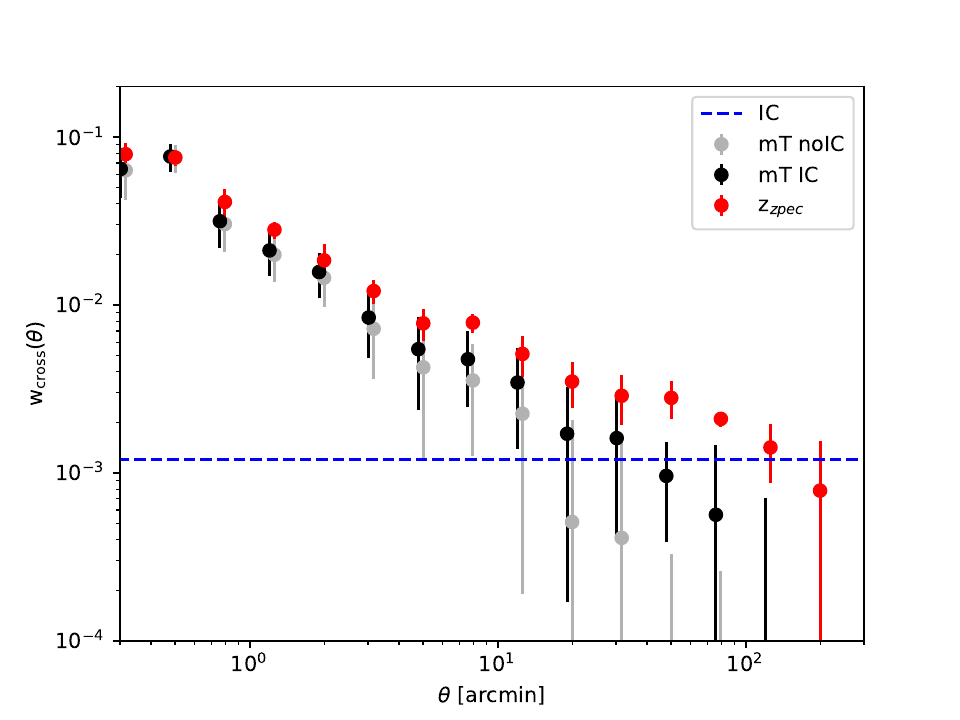}
 \caption{Comparison of the cross-correlation function estimated using different approaches. Grey circles correspond to the mini-tile approach before the IC correction is applied (see blue dashed line). Black circles correspond to the mini-tile results after the IC correction, while red circles are estimated using the new approach without any further correction. The uncertainties are derived from the covariance matrices.
 }
 \label{fig:mT}
\end{figure}

However, as illustrated in Fig. \ref{fig:mT}, the estimation of the IC correction is more complex than it may seem.
The cross-correlation function derived from a single measurement using the entire available area (shown as red circles), which can be considered unbiased with respect to the IC, differs from the mini-tiles one (black circles) even below 10 arcmin after the above IC value is applied. Therefore, if the IC is estimated using the measurements from the whole available area, the IC value would be larger due to the lower steepness of the cross-correlation function.

Moreover, the power-law approach is simply an effective approximation of the halo model \citep[see ][for a detailed description of the halo model used with magnification bias measurements]{Cue23}. If we estimate the IC assuming the theoretical halo model as the true model in Eq. (\ref{eq:ic}), but restricting to angular separations $<10$ arcmin, mostly unaffected by the cosmological model, it will underestimate the IC correction. On the other hand, if we include also the largest angular separations, the estimated IC correction can be affected by the cosmological model assumed. This means that the choice of angular separation limit becomes arbitrary, biased or model-dependent, which should be avoided.
Even more, the largest angular scales are the most important for constraining cosmological parameters, but they are also the most influenced by the adopted IC correction when using the mini-tile schema.

We attempted various methods to refine the IC value. Firstly, we conducted a polynomial fit analysis using the two sets of measurements, i.e., the mini-tiles, which were expected to be biased low, and the whole area measurements, which were likely biased high, as discussed later. The polynomial fit analysis yielded two extreme values for the IC: 7 and $18\times10^{-4}$. These limits defined the range of IC values that were compatible with our measurements.

To avoid the choice of an angular separation limit, we performed a power-law fit that included a constant value, the IC correction. However, most of the algorithms we tested were unable to converge to a solution. The derived IC value was $13\times10^{-4}$, which was in agreement with the previously defined range. Nevertheless, it was not possible to estimate a robust uncertainty for this value.

Next, we used theoretical estimates of the halo model to account for all angular scales but avoiding the power-law shape, specially at large angular scales where the model should decline. To restrict the astrophysical parameters and to be as independent as possible from the considered cosmological model, we followed this analysis. Firstly, we performed a Maximum Likelihood Estimation (MLE) search for the astrophysical parameters using only the 1-halo range measurements and random uniform cosmological parameters. With these astrophysical parameters, we obtained the IC distribution for another 100 random uniform cosmological parameter sets, resulting in a mean value of $15\times10^{-4}$ with a dispersion of $13\times10^{-4}$. However, several cosmologies were found to be incompatible with our data, so we further restricted them to those providing IC values in the range of $7-18\times10^{-4}$, resulting in a mean value of $11\times10^{-4}$ and a dispersion of $3\times10^{-4}$.

To test the sensitivity of our results to the assumed cosmological parameter distribution, we repeated the analysis using Gaussian random cosmological parameter sets based on the Planck ones \citep{PLANCK20} with a larger dispersion, $\sim0.05$. This resulted in a mean IC value of $13\times10^{-4}$ and a dispersion of $4\times10^{-4}$. Again, restricting the IC values to the range of $7-18\times10^{-4}$, we obtained a mean value of $13\times10^{-4}$ and a dispersion of $2\times10^{-4}$.

After considering the different cases, we selected a round number that fell between the most precise cases with a larger sigma. We choose a final IC value of $12\times10^{-4}$ with a dispersion of $3\times10^{-4}$ as the most appropriate IC estimate for our study using the mini-tile schema, taking into account the range of results obtained from various analyses and assumptions.

Figure \ref{fig:mT} provides a comparison of the cross-correlation function estimated using different approaches. Grey circles represent the mini-tile results before the IC correction, which show a sharp decline above 10 arcmin (where the blue dashed line, representing the IC correction, becomes relevant). Black circles correspond to the mini-tile results after applying the IC correction, while the red ones are estimated using the new approach without any further correction. The uncertainties are derived from the covariance matrices, which is described in the next section.

The mini-tile results are consistent with the new results within the uncertainties up to 30 arcmin, but they appear to be slightly underestimated across all angular scales. Above 30 arcmin, which are the most important for cosmological constraints, the new results are clearly higher than the mini-tile ones. However, the precise behaviour of the mini-tile results at these larger angular scales depends on the specific value of the IC correction, which also depends on the cosmological model. Therefore, such methodology should be avoided if there is a better alternative as the one proposed in this work. When considering higher IC values, such as IC=$15-20\times10^{-4}$, the two measurements become compatible even at the largest angular separations.

\section{Covariance estimation}
\label{sec:COV}

\subsection{Preliminaries on internal covariance estimation}


Let's assume we have a sample of size $N_s$ of $m$ independent random variables, $\{X_1,\ldots,X_m\}$. The covariance matrix can be estimated using the usual sample covariance formula, that is,

\begin{equation}
\text{Cov}(x_i,x_j) = \frac{1}{N_s-1} \sum_{k=1}^{N_s} (x_i^{k}-\bar{x}_i)(x_j^{k}-\bar{x}_j),
\end{equation}

where $x_i^{k}$ and $\bar{x}_i$ are the $k$-th observation and the sample mean of the variable $x_i$. The factor $\frac{1}{N_s-1}$ in the formula is called Bessel's correction and is used to correct for the bias in the sample covariance matrix due to estimating the population mean from the sample. 

This covariance matrix estimation is the natural choice when the measurement is obtained as the average of several independent observations and it is commonly known as ``subsampling''. However, it is clear that this approach cannot be used when only a single measurement is available, as it is our case using the whole available area. Moreover, according to \citet{NOR09}, the subsampling approach assumes that the subsamples used for estimating the covariance matrix are independent of each other, which is often not the case in galaxy clustering studies due to the presence of long-range modes in density fluctuations. This correlation between subsamples arises from the non-zero value of the correlation function at large scales. Therefore, alternative internal estimators need to be considered that can account for these limitations.

Jackknife and bootstrapping are alternative approaches based on ``resampling''. Jackknife is a leave-one-out method \citep{Sha86} where subsamples of the data are created by removing one observation at a time, resulting in $N$ subsamples with $N-1$ data points each. The covariance matrix is then estimated as:
\begin{equation}
    \text{Cov}(x_i,x_j)= \frac{N-1}{N} \sum_{k=1}^N \left( x_i^k-\bar{x}_i\right) \left( x_j^k-\bar{x}_j \right),
    \label{JKcov}
\end{equation}
where $x_i^k$ is the $k$-th Jackknife observation of the random variable $X_i$ and $\bar{x}_i$ is the mean of all Jackknife observations of the same random variable. The factor $\frac{N-1}{N}$ is included to take into account the fact that the resamplings of the data are not independent.

Bootstrap, on the other hand, involves randomly resampling the data $N_b$ times with replacement to create multiple subsamples of a given size $N$ \citep{Efr79}. The covariance matrix is then estimated as the sample covariance matrix of the subsamples, which can be written as:
\begin{equation}
    \text{Cov}(x_i,x_j)= \frac{1}{N_b} \sum_{k=1}^{N_b} \left( x_i^{k}-\bar{x}_i\right) \left( x_j^{k}-\bar{x}_j\right)\label{Bcov},
\end{equation}
where $x_i^k$ is the observation of the random variable $X_i$ coming from the $k$-th Bootstrap sample and $\bar{x}_i$ is the average over all $N_b$ Bootstrap samples.

Both Jackknife and Bootstrap are non-parametric approaches, meaning that they do not require any assumptions about the underlying distribution of the data. The choice between the two methods depends on the nature of the data and the research question at hand.

\subsection{Practical considerations}

In our case, the $m$ random variables described in the above section are the angular (auto-)cross-correlation function at different angular scales. However, one practical issue that arises when dealing with spatial data is that the naive method of resampling individual points (galaxies) does not work in this context. When subsampling spatial data, it is important to preserve the underlying dependence structure as much as possible in order to obtain valid estimates \citep{NOR09}. Thus, instead of resampling individual galaxies, the available area should be divided into sub-regions, and subsamples can be obtained by randomly selecting sub-regions rather than individual galaxies. This approach ensures that the subsamples retain the spatial dependence structure of the original data, which is crucial for the accurate estimation of the covariance matrix and other spatial statistics.

In essence, this means that, for the subsample method, we split the area into $N_s=63$ subregions (minitiles) and measure the auto/cross-correlation in each of those subregions, so that $x_i^k$ is interpreted, in this case, as the auto/cross-correlation measurement at angular scale $\theta_i$ from the subregion $k$, that is, $\tilde{w}^k(\theta_i)$. This is the method used in previous works \citep{BON20,BON21,GON21,CUE21,CUE22}. However, for the Jackknife and Bootstrap methods, we generate $N$ different subregions according to the previously described k-means algorithm and $x_i^k$ is to be interpeted as the auto/cross-correlation measurement at angular scale $\theta_i$ from either the $k$-th Jackknife subsample (that is, removing that $k$-th area) or the $k$-th Boostrap subsample.

For any subsampling method, to avoid a singular covariance matrix estimate, we need to ensure that the number of subsamples is sufficiently large relative to the number of measurements. The Johnson-Lindenstrauss lemma provides a mathematical justification for this requirement \citep{ACH03}. The mini-tiles schema, which uses $N_s$=63 subsamples versus 14 measurements, automatically satisfies this requirement. However, for other resampling methods, this implies that the natural subdivision into just the 4 fields may not be sufficient, hence the need for a k-means clustering algorithm to define patches in each of the fields.

At the same time, it is important to balance the size of the sub-regions with their representativeness of the dataset. Larger sub-regions may decrease the correlation between sub-regions and the influence of pairs crossing between sub-regions. However, increasing the size of the sub-regions reduces the number of subsamples that can be obtained, which increases the variance of the covariance estimator and may require a larger number of subsamples to obtain a reliable estimate \citep{Fri16}.

Finally, it should be noted that internal covariance estimation is valid only in the limit where the correlations between sub-regions are small. In this case, for instance, the sub-sample covariance matrix becomes a rescaling of the sample covariance  matrix of independent realisations of the sub-regions. The internal estimate for the inverse of the covariance matrix will then be proportional to the true covariance matrix \citep{Fri16}. As a rule of thumb, the sub-regions should be large enough so that their effective angular scale is well above the maximum measured scaled. A full treatment of covariance estimation would involve a study using cosmological simulations tailored for magnification bias, which is beyond the
scope of this paper but will be addressed in future work.

\subsection{Internal covariance matrix estimations}

\begin{figure*}[ht]
\includegraphics[width=\textwidth]{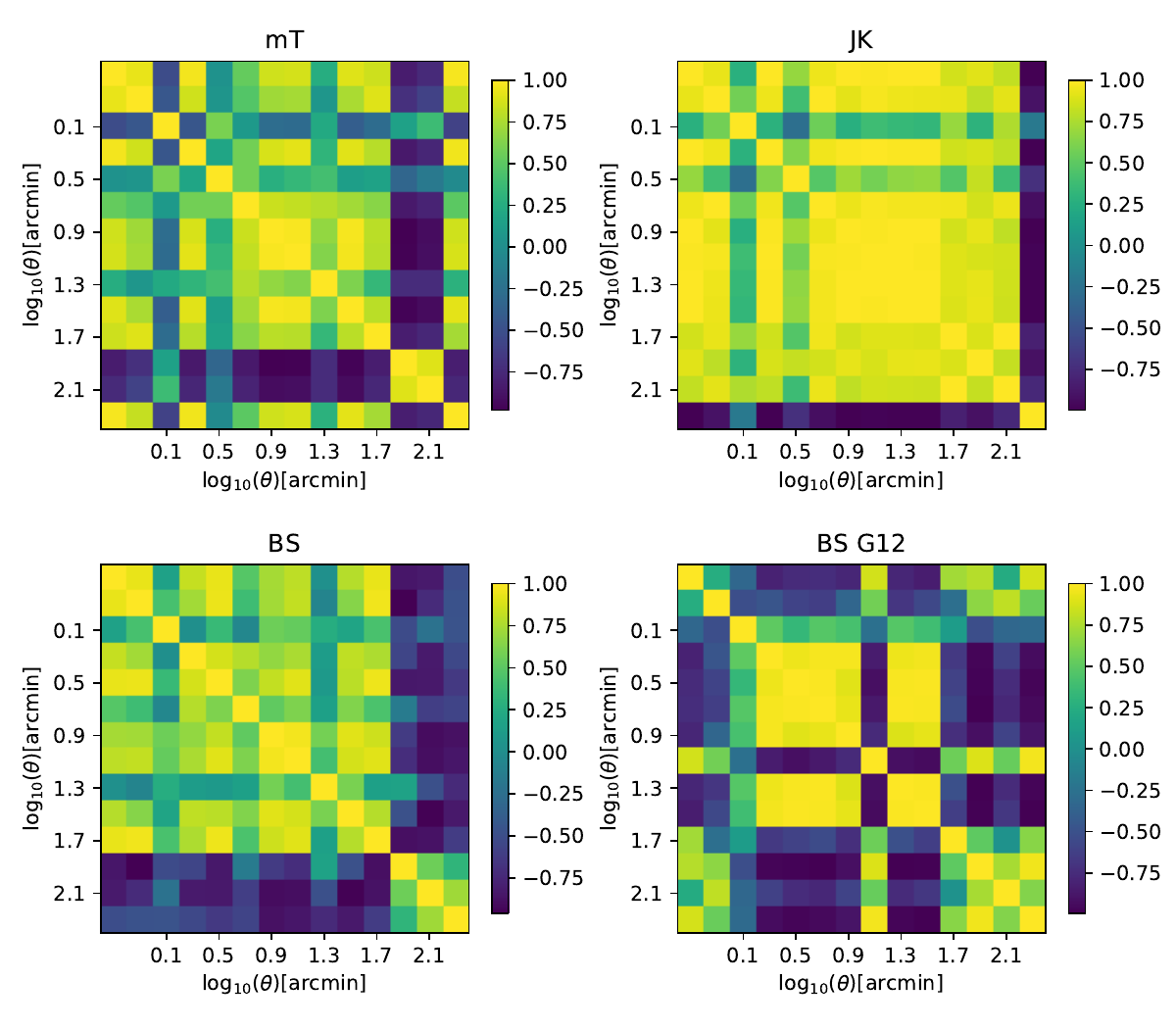}
 \caption{Comparison of the correlation matrices estimated with three methods for the cross-correlation signal. The mini-tiles estimation (top left, mT) and the bootstrapping estimation (bottom left, BS) are similar, with higher correlation at central angular distances. However, the Jackknife estimation (top right, JK) shows strong correlation across all angular separations, indicating an implementation issue. To address potential signal variation among fields, the covariance matrix is estimated for each individual field using 16 patches of approximately three square degrees each (bottom panel, BS G12; example for the G12 field).
 }
 \label{fig:COV}
\end{figure*}

In light of the practical considerations discussed earlier, we chose to divide the fields into patches of approximately nine square degrees, resulting in a total of $N$=22 patches, as described in Sect. \ref{sec:tiling}. The bottom panel of Fig. \ref{fig:patches5} shows the five patches defined for the G15 field.

For our leave-one-out Jackknife method implementation, we remove one patch from the data and estimate the cross-correlation with the remaining area in each iteration. This approach minimises potential effects on the spatial dependence structure of the original data, since we do not modify three of the fields in each iteration. Furthermore, this speeds up the calculations as the number of pairs for each field can be stored and reused between iterations. In essence, we have $N=22$ Jackknife subsamples, which we plug into \eqref{JKcov} to compute the associated covariance matrix.

In our Bootstrap implementation, we generate bootstrapped samples by randomly selecting, with repetition, a certain number of patches between the available ones. While the typical number of patches should be $N$, that is, 22 in our case, following \citet{NOR09}, we over-subsampled by a factor of 3, resulting in a total of 66 patches in this case. We confirmed their conclusion that the estimated covariance matrix using over-subsampling is more robust. We generated $N_b$=10000 bootstrapped subsamples and found that increasing this number did not yield any significant improvement in the covariance matrix estimation, computed by plugging the measurements derived from the subsamples into \eqref{Bcov}.

For ease of comparison, we chose to display the correlation matrices instead of the covariance matrices since they are directly related. 
The correlation matrices estimated with the three methods are compared in Figure \ref{fig:COV}. The mini-tiles estimation (top left) is very similar to the bootstrapping estimation (bottom left), but noisier. In both cases, the central angular distances (in arcmin and on a logarithmic scale) show higher correlation than the rest, corresponding to the transition between the 1- and 2-halo regimes. For the mini-tiles estimation, we added the uncertainty of the IC correction in quadrature to the diagonal of the covariance matrix.

However, the Jackknife estimation (top right) is markedly different from the other methods. It shows a strong correlation between all angular separations, indicating a problem with the implementation. As we will discuss in the next section, there exists a signal variation between the different fields that violates the validity condition for the internal covariance estimation in this particular implementation. The other estimations are less affected by this issue due to the fact that they are based on smaller and more homogeneous regions. Considering that we prefer to avoid the mini-tile approach because of its strong dependence on the IC value and the worse performance of the Jackknife estimation, we adopted the bootstrapping one as our default approach to estimate the covariance matrices.

To study the potential variation of the cross-correlation signal among the different fields, we need to estimate the covariance matrix in each of them individually. In those cases, we increased the number of patches per field to 16, which provides an area of approximately three square degrees per patch, even smaller than the mini-tiles. The bottom panel of Fig. \ref{fig:patches16} displays the 16 patches defined for the G09 field. It should be noticed that the IC correction is not an issue for the covariance estimation for either the mini-tiles or the Bootstrap method because it is cancelled in the calculation. 

In the case of individual fields, both the bootstrapping and Jackknife methods produced similar results due to the higher homogeneity between the patches. To maintain consistency, we adopted the bootstrapping method as our default to estimate the covariance matrices in all cases. Therefore, the error bars displayed in Figure \ref{fig:mT} are simply the square root of the diagonal terms of the covariance matrices for the mini-tiles and bootstrapping methods, respectively.

\section{Robustness and sample variance}
\label{sec:robust}
\begin{figure}[ht]
\includegraphics[width=\columnwidth]{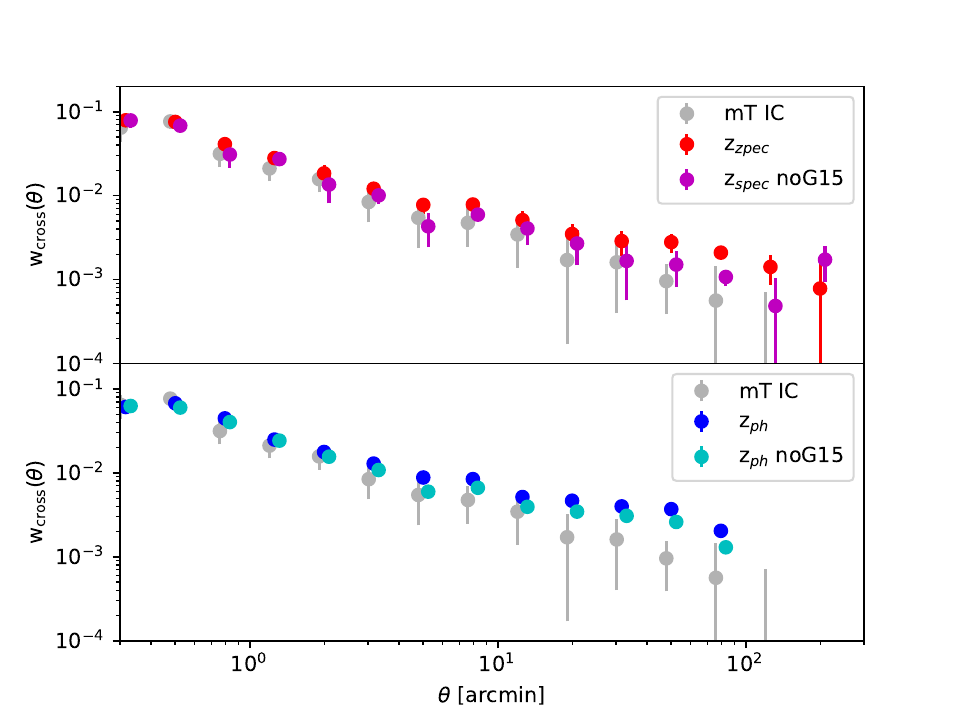}
 \caption{Comparison of the cross-correlation function``whole-area'' measurements from the z$_{spec}$ and z$_{ph}$ lens samples with the IC-corrected mini-tiles measurements (common in both panels, gray circles). The top panel shows the measurements from the z$_{spec}$ sample (red circles) and without the G15 field (magenta circles), while the bottom panel shows the measurements from the z$_{ph}$ sample (blue circles) and without the G15 field (cyan circles).
 }
 \label{fig:zspec_vs_zph}
\end{figure}

We investigate the robustness of our results by applying the new methodology to a new lens sample obtained from the photometric SDSS catalogue (described in Sect. \ref{sec:data}), and comparing the resulting cross-correlation function with that measured from the z$_{spec}$ sample. The z$_{spec}$ sample covers the G09, G12, G15, and SGP regions, while the z$_{ph}$ sample excludes SGP and includes the NGP region instead. The measurements from the z$_{spec}$ sample (red circles, top panel) and the z$_{ph}$ one (blue circles, bottom panel) are plotted against the IC-corrected mini-tiles measurements in Fig. \ref{fig:zspec_vs_zph}. In line with the findings of \citet{GON21}, both samples produce compatible measurements, but the z$_{ph}$ measurements are slightly higher at 30-40 arcmin and do not exhibit statistically significant cross-correlation values above 100 arcmin. However, the two measurements are clearly above the mini-tiles ones for angular separations greater than 10 arcmin, as previously mentioned. The excess using different tiling approaches with respect the mini-tiles case was interpreted as a large-scale bias by \citet{GON21}. It is important to note that the large-scale behaviour of the mini-tile measurements depends strongly on the IC correction applied, which in turn depends on the assumed cosmological parameters. 

Furthermore, the GAMA sample used in our study comes from a spectroscopic survey, which is fundamentally different from our photometric SDSS sample. Therefore, any large-scale discrepancies between both measurement approaches cannot be attributed to systematic errors related to the survey characteristics. Instead, it suggests that the differences may be due to physical properties of the samples themselves, such as the distribution of foreground or background galaxies, in the regions being probed. We will further investigate the nature of this discrepancy to determine whether it is a bias or a consequence of spatial variance within the samples. 

\subsection{Sample variance} 
\begin{figure*}[ht]
\includegraphics[width=\textwidth]{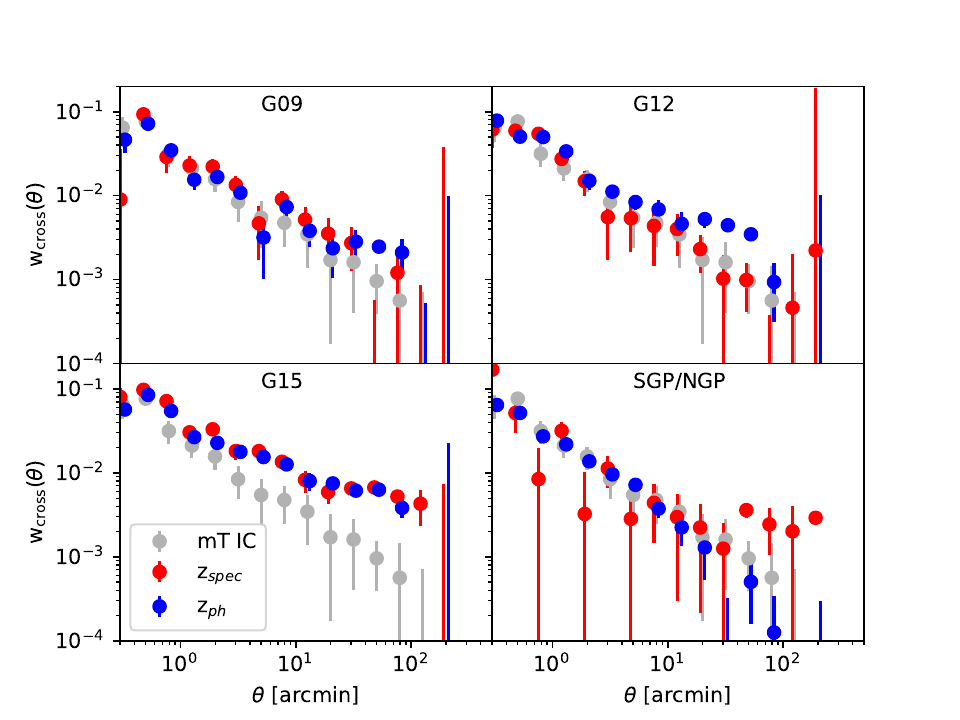}
 \caption{Comparison of the cross-correlation functions estimated for each individual field for the spectroscopic sample (z$_{spec}$; red circles) and the photometric one (z$_{ph}$; blue circles). The mini-tiles measurements for the whole area are the
same in each panel and are added as a visual guide (gray circles).
 }
 \label{fig:xc_fields}
\end{figure*}

We estimated the cross-correlation function for each individual field, and the uncertainties were derived using a bootstrap covariance matrix estimated using 16 patches per field. The results for each sample are compared in Fig. \ref{fig:xc_fields}, where the red circles represent the z$_{spec}$ sample and the blue circles represent the z$_{ph}$ sample. The mini-tiles measurements for the whole area using the z$_{spec}$ sample are the same in each panel and are added as a visual guide.

Focusing on the z$_{spec}$ sample results, we observed that the measurements for the G09, G12, and SGP fields were mostly compatible with the mini-tile measurements. However, the G15 field showed a stronger signal compared to the other fields at almost all angular separation distances. Although the SGP field showed a hint of a stronger signal above 50 arcmin, it had the worst stability and errors. 

Similar conclusions were drawn from the z$_{ph}$ sample results. We observed a stronger signal above 30 arcmin for the G09 and G12 fields, which was not compensated for by the slightly lower signal for the NGP field. In contrast to the z$_{spec}$ results, no positive correlation was observed above 100 arcmin in any field with the z$_{ph}$ sample. These data points showed a rebound effect and, in general, had larger error bars.


Therefore, the observed stronger signal on large scales, compared to the mini-tiles case, was primarily produced by the G15 region in both samples, while it was also present in the G09 and G12 fields of the z$_{ph}$ one, albeit to a lesser extent. These findings suggest that the discrepancy is not indicative of a general large-scale bias caused by the sample or methodology, as interpreted by \citet{GON21}, but rather it may be attributed to a sample variance issue related to the large-scale structure in the G15 region. Notably, when the G15 region is excluded from the cross-correlation estimation (see Fig. \ref{fig:zspec_vs_zph}), the discrepancy is mitigated for the z$_{spec}$ sample, but only alleviated for the z$_{ph}$ sample.

Several studies have investigated the scale of isotropy in the galaxy distribution. \citet{Mar12} analyzed luminous red galaxies from the SDSS DR7 and found a scale of isotropy at $150\,h^{-1}$ Mpc, while \citet{Sar19} indicated a transition to isotropy at a scale of approximately $200\,h^{-1}$ Mpc for both SDSS photometric and spectroscopic data. These findings support the notion that the galaxy distribution in the local Universe is isotropic on scales larger than $200\,h^{-1}$ Mpc, affirming the validity of assuming isotropy on large scales. However, considering that the typical physical scales of our fields are below $200\,h^{-1}$ Mpc at the mean redshift of the samples, it is expected that sample variance plays a role in the observed discrepancies.

\begin{figure*}[ht]
\includegraphics[width=\textwidth]{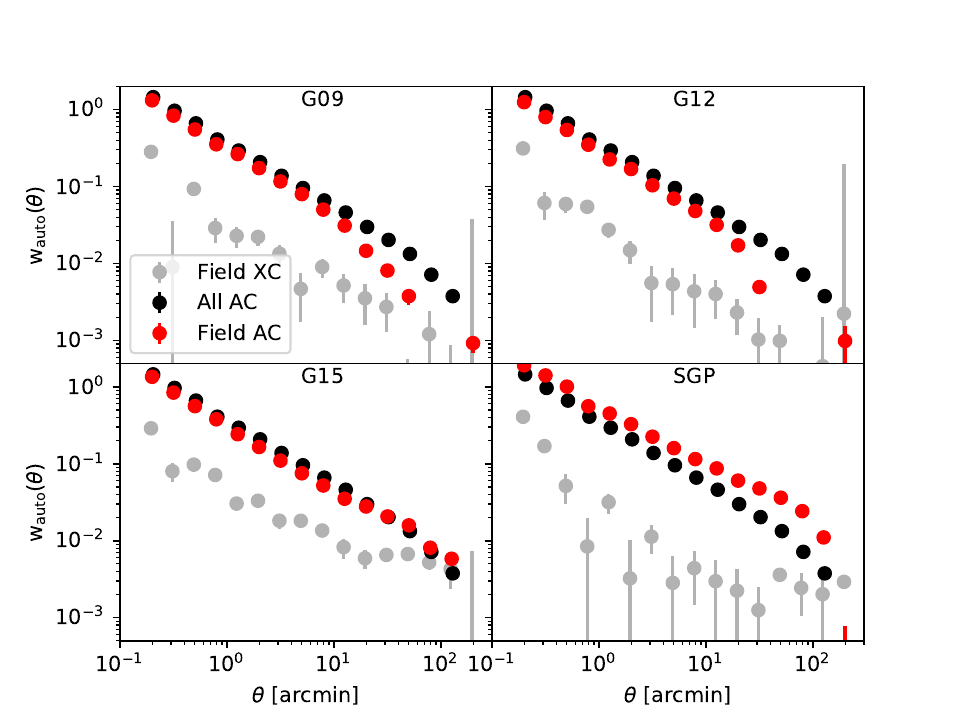}
 \caption{The auto-correlation functions for the z$_{spec}$ foreground sample estimated for all the fields (black circles) are compared to the ones derived from individual fields (red circles) in this figure. The cross-correlation functions estimated for each individual field are also shown for comparison (grey circles).
 }
 \label{fig:ac_fg}
\end{figure*}

Sample variance is the statistical uncertainty that arises because measurements are made in a finite volume (or on a finite sky area), so only a limited number of independent modes or regions are observed.
In order to test this hypothesis, we study the statistical properties of the samples, both foreground and background. A visual comparison between the galaxy distribution of G15 (see Fig. \ref{fig:patches5}) and G09 (see Fig. \ref{fig:patches16}) shows no significant difference: for the foreground sample (top panels) both regions show an equivalent amount of large-scale structure, and for the background ones, the only interesting aspect is the overlapping regions that are common in all the fields and are already taken into account in the random catalogues.

\begin{figure*}[ht]
\includegraphics[width=\textwidth]{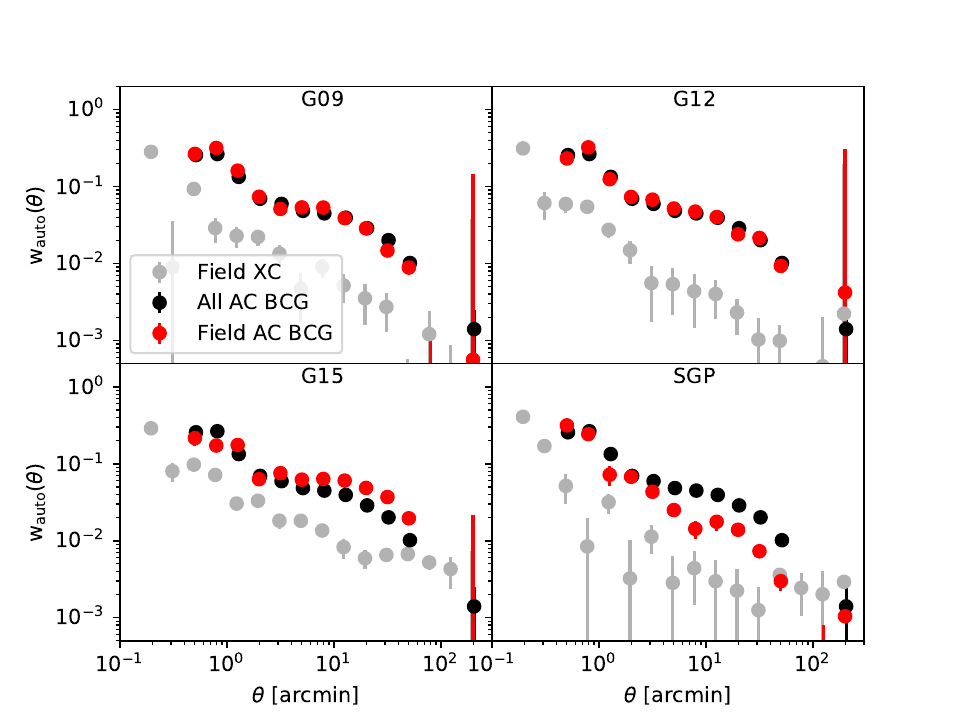}
 \caption{The auto-correlation functions for the background sample estimated for all the fields (black circles) are compared to the ones derived from individual fields (red circles) in this figure. The cross-correlation functions estimated for each individual field are also shown for comparison (grey circles).
 }
 \label{fig:ac_bk}
\end{figure*}

For a more quantitative analysis, we estimate the angular auto-correlation function for all the fields (black circles) and compare them to the ones derived from individual fields (red circles). We repeat the same analysis for both the foreground (see Fig. \ref{fig:ac_fg}) and background samples (see Fig. \ref{fig:ac_bk}). The cross-correlation function estimated for each individual field is also plotted as a comparison (grey circles).

As expected, the auto-correlation function from the whole area displayed the typical smooth shape with no apparent transition between the 1- and 2-halo regimes and is stronger than the cross-correlation function. We compared the auto-correlation functions estimated from each individual field and found that the regions had compatible auto-correlation values for separations below 10 arcmin, with the SGP region showing a slightly stronger correlation. However, the behaviour of the auto-correlation function for large scales varied between the regions. 

For instance, the auto-correlation function for G15 maintained a linear behaviour (in logarithmic scale) until it became negative beyond 100 arcmin, while the auto-correlation function for G09 and G12 clearly declined before such an angular separation. This result is interesting because the overall projected correlation function measured for the same GAMA sample \citep{Sur21} and the most general SDSS one \citep{Zeh05} show no decline at least until 200-300 arcmin, suggesting that G09 and G12 are regions with less large-scale structure compared to G15, which is not anomalous in this sense.

On the other hand, the SGP region displayed a clear excess compared to the other regions, indicating an excess of large-scale structure in the SGP and G15 fields compared to the G09 and G12 fields. This suggests that the particular behaviour of our lensing results in the G15 field is likely due to the presence of the stronger large-scale structure. However, the cross-correlation function for the SGP region was much weaker than for G15, indicating that this excess alone cannot explain the observed lensing results.

In the case of the background sample, the auto-correlation functions derived from the different regions are in relatively good agreement.  The auto-correlation functions are estimated above 0.4 arcmin due to the beam size of Herschel detectors, and they become negative above 50 arcmin due to the angular distance dependence with redshift. 

Overall, the background auto-correlation function for the G15 region is very similar to those of the G09 and G12 regions, showing only slightly higher correlation. In contrast, the SGP region shows a much weaker auto-correlation function. This suggests that the anomalous behaviour of G15 may be explained by a combination of a stronger large scale structure in both the foreground and background samples. It is important to note that this is in contrast to the SGP region, which shows a stronger signal in the foreground sample but a much weaker one in the background sample. 

\section{Conclusions}
\label{sec:concl}
This study provides a comprehensive analysis of a refined methodology for submillimeter galaxy magnification bias measurements, focusing on the angular cross-correlation function and associated uncertainties. The subsampling approach based on mini-tiles has been widely used to estimate the cross-correlation function between galaxy samples thanks to its computational efficiency. However, large-scale fluctuations can introduce a bias into the estimated cross-correlation function, and an IC term is required to obtain an unbiased estimate of the true function. This correction is particularly crucial for constraining cosmological parameters on the largest angular scales. However, accurately determining the value of the IC correction can be challenging due to its dependence on factors such as the survey geometry, observational properties of the galaxy samples, and the cosmological model. In light of these challenges, we have adopted a new methodology to obtain more robust measurements for constraining cosmological parameters.

The new methodology to estimate the cross-correlation function involves a statistically rigorous approach that uses the full field area to count the number of different pairs for each field and combine them into a single estimation. This method reduces statistical uncertainty and accounts for the full information available in the data, in contrast to measuring separate (auto-)cross-correlation functions for each field or subregion in a field. The statistical superiority of this approach over the mini-tile and zone-averaging alternatives is further demonstrated through simulations in Appendix \ref{secA1}. To estimate the covariance matrix we adopted an oversampled bootstrap method by dividing each field in at least five patches, the minimum number to maximise the patches area but maintaining a number greater than the number of measurements. The patches were defined automatically using a k-mean clustering algorithm. This resampling approach ensures that the subsamples retain the spatial dependence structure of the original data. For analysing individual fields we divided them in 16 patches, instead. The new measurements were demonstrated to be more robust and with much lower uncertainties. A comprehensive treatment of covariance estimation would require a study using cosmological simulations specifically designed to investigate magnification bias. This topic is beyond the scope of this paper but will be addressed in future work.

We investigate the robustness of the new methodology for measuring the cross-correlation function by comparing the new results from a spectroscopic lens sample with those from a photometric lens sample. The consistency between the two independent foreground samples, which have completely different systematic properties, provides empirical support that the large-scale signal recovered by the new method is not an artefact of the estimator. Moreover, it also suggests that the differences with respect the mini-tile approach may be due to the complexity of computing the integral constraint in a consistent manner, but also to the physical properties of the samples themselves given the behaviour of the large-scale fall.

We further analyse the cross-correlation function and auto-correlation function for individual fields in the GAMA regions, comparing the z$_{spec}$ and z$_{ph}$ samples. The G15 field was found to have a stronger signal compared to the other fields at almost all angular separation distances for both the z$_{spec}$ and z$_{ph}$ samples. The results suggest that there is no general large-scale bias produced by the sample or methodology, but rather, it may be more related to a sample variance problem of the large scale structure at the G15 region. In fact, the SGP and G15 regions showed excess large-scale structures in the foreground sample compared to the G09 and G12 regions. However, this excess alone could not explain the observed lensing results because the SGP showed a weaker cross-correlation signal. Considering the background sample, the auto-correlation function for the G15 region was very similar to those of the G09 and G12 regions, showing only slightly higher correlation, while the SGP region showed a much weaker auto-correlation function. Therefore, it seems that the stronger cross-correlation found in G15 is produce by the rare combination of two excess of large-scale structure in both the foreground and background samples.

Despite the stronger signal, there is currently no indication that the G15 region should be discarded, and with the addition of more independent regions in future analyses, it will be possible to perform a statistical analysis to quantify the degree of anomaly in the G15 results and its cosmological implications.

\backmatter

\bmhead{Acknowledgements}

JGN, LB, MC, DC, JMC and RF acknowledge the PID2021-125630NB-I00 project funded by MCIN/AEI/10.13039/501100011033/FEDER, UE. \\

The Herschel-ATLAS is a project with Herschel, which is an ESA space observatory with science instruments provided by European-led Principal Investigator consortia and with important participation from NASA. The H-ATLAS web- site is http://www.h-atlas.org. GAMA is a joint European- Australasian project based around a spectroscopic campaign using the Anglo- Australian Telescope. The GAMA input catalogue is based on data taken from the Sloan Digital Sky Survey and the UKIRT Infrared Deep Sky Survey. Complementary imaging of the GAMA regions is being obtained by a number of independent survey programs including GALEX MIS, VST KIDS, VISTA VIKING, WISE, Herschel-ATLAS, GMRT and ASKAP providing UV to radio coverage. GAMA is funded by the STFC (UK), the ARC (Australia), the AAO, and the participating institutions. The GAMA web- site is: http://www.gama-survey.org/.\\
Funding for the Sloan Digital Sky Survey IV has been provided by the Alfred P. Sloan Foundation, the U.S. Department of Energy Office of Science, and the Participating Institutions. SDSS-IV acknowledges
support and resources from the Center for High-Performance Computing at
the University of Utah. The SDSS web site is www.sdss.org.

SDSS-IV is managed by the Astrophysical Research Consortium for the 
Participating Institutions of the SDSS Collaboration including the 
Brazilian Participation Group, the Carnegie Institution for Science, 
Carnegie Mellon University, the Chilean Participation Group, the French Participation Group, Harvard-Smithsonian Center for Astrophysics, 
Instituto de Astrof\'isica de Canarias, The Johns Hopkins University, Kavli Institute for the Physics and Mathematics of the Universe (IPMU) / 
University of Tokyo, the Korean Participation Group, Lawrence Berkeley National Laboratory, 
Leibniz Institut f\"ur Astrophysik Potsdam (AIP),  
Max-Planck-Institut f\"ur Astronomie (MPIA Heidelberg), 
Max-Planck-Institut f\"ur Astrophysik (MPA Garching), 
Max-Planck-Institut f\"ur Extraterrestrische Physik (MPE), 
National Astronomical Observatories of China, New Mexico State University, 
New York University, University of Notre Dame, 
Observat\'ario Nacional / MCTI, The Ohio State University, 
Pennsylvania State University, Shanghai Astronomical Observatory, 
United Kingdom Participation Group,
Universidad Nacional Aut\'onoma de M\'exico, University of Arizona, 
University of Colorado Boulder, University of Oxford, University of Portsmouth, 
University of Utah, University of Virginia, University of Washington, University of Wisconsin, 
Vanderbilt University, and Yale University. \\

This research has made use of the python packages \texttt{ipython} \citep{ipython}, \texttt{matplotlib} \citep{matplotlib} and \texttt{Scipy} \citep{scipy}.

This version of the article has been accepted for publication, after peer review but is not the Version of Record and does not reflect post-acceptance improvements, or any corrections. The Version of Record is available online at: https://doi.org/10.1007/s10686-026-10055-x

\section*{Declarations}

\begin{itemize}
\item Funding This work was supported by the PID2021-125630NB-I00 project funded by MCIN/AEI/10.13039/501100011033/FEDER, UE.
\item Competing interests The authors declare no competing interests.
\item Data availability The data used in the study is publicly available.
\item Author contribution All authors contributed to the study conception and design. The first draft of the manuscript was written by J. González-Nuevo and all authors commented on previous versions of the manuscript. All authors read and approved the final manuscript.
\end{itemize}

\begin{appendices}

\section{Statistical characterization of the different cross-correlation function estimation approaches}\label{secA1}
Identifying the best measurement method for a particular dataset is generally a difficult task that requires access to highly realistic simulations, which are not always available or even fully developed. The alternative is to rely on robust, well-tested methodologies. The objective of this appendix is to characterize the statistical properties of the different approaches discussed in this work and determine which is the most robust.

We compare three cross-correlation function measurement approaches: a single estimation using the full available area, the average of the estimations for each H-ATLAS field individually (N = 4), and the average of estimations from the more numerous but smaller mini-tiles (N = 64). We focus on a single angular separation to simplify the analysis, adopting a fixed value of w = 0.01; the exact value is not relevant to our conclusions.

The mini-tiles are assigned random areas between 10 and 20 deg$^2$, and the number of random pairs, RR, is rescaled accordingly relative to the smallest area as a reference. Adopting the \citet{DP83} estimator, the DD values are calculated as $DD = (1 + w) * RR$. We then introduce a Gaussian fluctuation around the DD values assuming different increasing dispersion levels, $\sigma(DD)$. Finally, the cross-correlation is calculated for each of the three approaches: the average of the 64 mini-tiles values, $<\hat{w}_{64}>$; the average of the four zone values, $<\hat{w}_4>$, each estimated as $\hat{w} = \sum{DD}/\sum{RR} - 1$ considering all the pairs in each zone, where each zone contains approximately 16 mini-tiles; and the single value $\hat{w}_1$ from the full-area approach, combining pairs from all mini-tiles. The uncertainties are calculated as $\hat{\sigma}_w = STD(\hat{w}_i)/\sqrt{N}$ for the mini-tiles and zones approaches, and as $\sigma(DD)/\sum(RR)$ for the full-area estimator.

\begin{figure*}[ht]
\includegraphics[width=\textwidth]{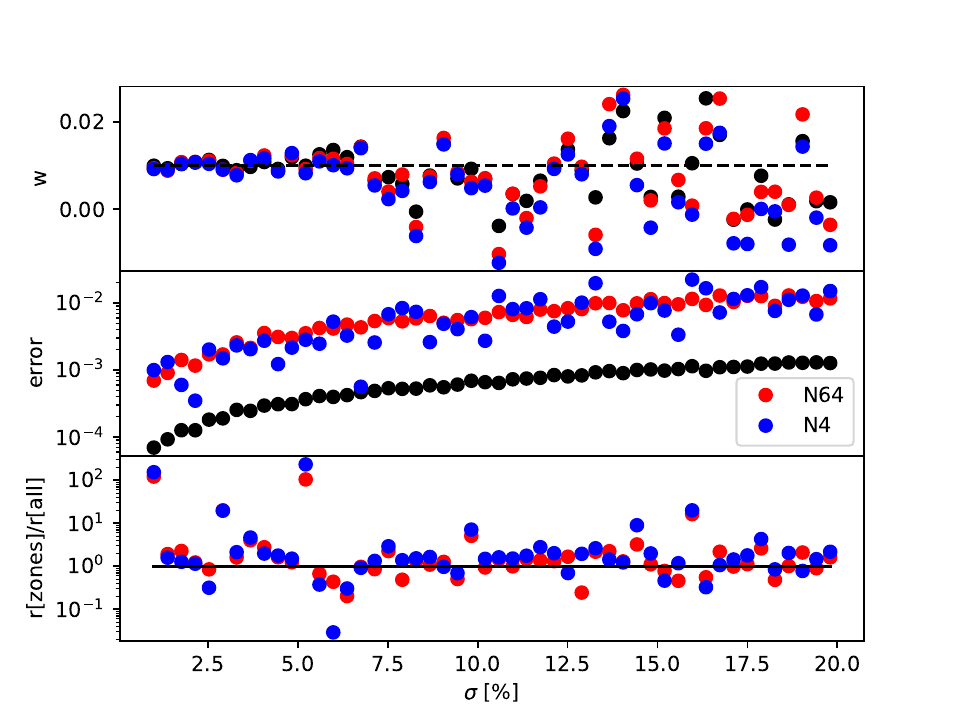}
 \caption{Statistical characterization of the cross-correlation at a fixed angular separation (or $\theta$) for different estimation approaches assuming a potential dispersion in the measurements (number of data-data pairs): 64 independent mini-tiles (red), 4 independent zones of approximately 16 mini-tiles each (blue) and using all the available area at the same time (black). The estimated cross-correlation and its uncertainty are shown in the top and middle panels, respectively. The comparison between the relative error with respect the all-area case is shown in the bottom panel.
 }
 \label{fig:bias_N64}
\end{figure*}

Figure \ref{fig:bias_N64} summarizes the statistical characterization of the three approaches derived from this simulation setup. The top panel shows the estimated cross-correlation for each approach at fixed fluctuation level (mini-tiles in red, 4 zones in blue, full area in black). At low fluctuation levels, representative of small angular separations, all approaches yield similar measurements close to the input value. At higher fluctuation levels, representative of large-scale regimes, the estimated correlations fluctuate around the input value, but the full-area approach consistently shows the smallest dispersion.

The middle panel compares the estimated uncertainties for each approach. Because the mini-tile and 4-zone approaches rely on the standard deviation between their respective sub-measurements, their uncertainties are systematically larger than those from the full-area estimator. The uncertainties from the 4-zone approach fluctuate more than those from the mini-tile approach, making it less reliable in this respect. Including a full covariance matrix treatment for the uncertainties would not change the conclusions of this comparison.

The bottom panel shows the ratio of the relative errors of the mini-tile and zone approaches to that of the full-area case, serving as a proxy for the bias comparison between approaches. While individual simulations occasionally produce less biased results for the subdivided approaches compared to the full-area one, the median (mean) ratios across all simulations are 1.30 (6.6) for the mini-tile approach and 1.52 (10.2) for the 4-zone approach. These results are consistent with the empirical conclusions discussed in this work and in \citet{GON21}.

Therefore, this analysis confirms the well-established result that using the full available area yields more robust measurements with smaller uncertainties. Although particular random realizations may exist in which subdividing the area produces a better individual measurement, this cannot be predicted in advance for real data. 

We acknowledge that this simulation setup is intentionally simplified and does not reproduce the full complexity of the observational data, including realistic survey geometry, redshift distributions, and scale-dependent signal behaviour. A complete validation using realistic mock catalogues tailored to the magnification bias setup is planned for future work. Nevertheless, the results presented here provide a useful statistical characterization of the relative performance of the three approaches under controlled conditions.

\end{appendices}

\noindent

\bibliography{magbias_method}

\end{document}